\documentclass[journal=jctcce,manuscript=article]{achemso}

\usepackage{amsmath}
\usepackage{amssymb}
\usepackage{booktabs}
\usepackage{graphicx}
\usepackage{xcolor}
\usepackage{bm}

\newcommand*{\pprpa}{ppRPA}

\title{Development of a Minimal Auxiliary Basis for the Particle--Particle Random Phase Approximation (ppRPA): Accelerating Excited-State Calculations in Diradical Nanoparticles and Periodic Systems}

\author{Yuchen Wang}
\affiliation{State Key Laboratory of Precision and Intelligent Chemistry, University of Science and Technology of China, Hefei, China}
\email{yuchenwang2015@ustc.edu.cn}

\author{Zhenyu Li}
\affiliation{State Key Laboratory of Precision and Intelligent Chemistry, University of Science and Technology of China, Hefei, China}
\email{zyli@ustc.edu.cn}

\begin{document}

\begin{abstract}
Singlet--triplet gaps set the energy scale of the two-spin manifold in
molecular quantum information materials, and such systems frequently carry
diradical character: metal nanoclusters, metal--organic frameworks with radical
linkers, and the nitrogen--vacancy center in diamond.  Time-dependent density
functional theory is the standard tool at this size but describes
multireference systems poorly.  A closed-shell reference is inadequate for a
multiconfigurational ground state, and the adiabatic kernel excludes doubly
excited configurations.  The particle--particle random phase approximation
(\pprpa{}) changes the particle number rather than the spin.  Singlet and
triplet states descend from one closed-shell reference, and double excitations
enter on the same footing as single ones.  Computational cost has confined
\pprpa{} to small systems, and the two-electron integrals that drive it are
factorized through an auxiliary basis.  We develop \pprpa{}--as, which reduces
that basis to a minimal set of atom-centered Gaussian shells whose exponents
follow from free-atom radii and one global scaling factor.  The approximation
enters only the excited-state step, so both calculations proceed from the same
self-consistent reference.  We test molecules and clusters of up to 286 atoms,
including heavy elements, and seven point defects in four periodic hosts,
obtaining both multiplicities and their splitting without symmetry breaking or
spin projection.  We optimize the scaling factor and give guidance on choosing
it.  An order-of-magnitude smaller auxiliary basis delivers a speed-up of up to
eightfold while staying within $0.03$~eV of conventional \pprpa{}, bringing
diradical nanoclusters and periodic systems within reach.
\end{abstract}

\section{Introduction}
\label{sec:intro}

The relative energies of the lowest singlet and triplet states determine
whether a molecular material works.  In thermally activated delayed
fluorescence, a gap small enough to permit thermally driven reverse
intersystem crossing lets an organic light-emitting diode harvest triplet
excitons that would otherwise be lost \cite{Uoyama2012Highly}.  In singlet
fission, one photogenerated singlet becomes two triplets through a correlated
triplet-pair state of doubly excited character \cite{Smith2010Singlet}.  In
molecular quantum information science, the exchange coupling between two
weakly interacting unpaired electrons sets the energy scale of the two-spin
manifold, and hence the addressability of the qubit
\cite{GaitaArino2019Molecular,Wasielewski2020Exploiting}.  The deciding
quantity throughout is the separation between the lowest singlet and triplet
arising from two electrons in a pair of near-degenerate frontier orbitals.
The high-spin triplet is a single determinant and is reproduced by almost any
single-reference method.  The singlet is not: it mixes the closed-shell
occupations $\varphi_1^2$ and $\varphi_2^2$ with the open-shell pairing, and
its position depends on the exchange integral $K_{12}$ and on the
configuration mixing that $\Delta\varepsilon_{12}$ controls.  An unbalanced
description propagates directly into the splitting, which is small enough that
the error is often as large as the quantity itself \cite{Slipchenko2002Singlet,Bernard2012General,Casanova2020SpinFlip,%
Penfold2015Predicting}.

Multireference wavefunction methods describe the two multiplicities evenly, but
their cost grows rapidly with the active space, which confines them to systems
smaller than those of interest here.  Shrinking the system helps less than it
appears to: the ligand shell that provides the chemical tuning accounts for
most of the atoms, and it can be simplified but not removed.  Single-reference
methods are affordable at this size, but their accuracy depends strongly on the
reference determinant.  A practical method must therefore treat both
multiplicities evenly, accommodate the configuration mixing the singlet
requires, and remain usable at a few hundred atoms.

Linear-response time-dependent density functional theory (TDDFT) is the
workhorse for excited states at this size
\cite{Runge1984Density,Casida1995Time}.  Its limitations follow from its
reference.  TDDFT is formulated about the $N$-electron ground state and
propagates particle--hole amplitudes \cite{Marques2004TimeDependent}.  The
electron number is therefore fixed, and the response space spans single
excitations only.  Within the adiabatic approximation the
exchange--correlation kernel is frequency independent, so states with
substantial doubly excited character are absent from the spectrum
\cite{Maitra2004Double,Dreuw2005Single}.  The same omission makes the topology
of conical intersections with the ground state qualitatively wrong
\cite{Levine2006Conical}.  For diradicals the problem is more basic.  The
ground state is itself multiconfigurational, so a closed-shell reference is
already inadequate, and the two multiplicities are not treated evenly.
Spin-flip methods address that imbalance by choosing a high-spin reference
\cite{Krylov2001Size}.  The limiting approximation for this class of problem
is therefore the reference determinant rather than the functional applied to
it \cite{Casanova2020SpinFlip}.

The particle--particle random phase approximation (\pprpa{}) then changes another variable: it changes the
particle number instead of the spin
\cite{VanAggelen2013Exchange,VanAggelen2014Exchange,Yang2013Benchmark}. One eigenvalue problem carries two branches. The particle--particle branch adds a pair of electrons to an $N-2$ reference, and the hole--hole branch removes a pair from an $N+2$ reference, so the same neutral target can be reached from either side \cite{Yang2013Double,Bannwarth2020Hole}. The $N$-electron states follow by adding or removing an electron
pair.  This leads to three things.  First, every placement of the pair is an
eigenvector.  Configurations that are doubly excited with respect to the
$N$-electron ground state therefore enter on the same footing as singly excited
ones.  Second, the whole low-lying $N$-electron manifold comes from one
diagonalization with ground state included.  Excitation energies are differences
between its eigenvalues, so the reference energy cancels exactly.  Third,
singlet and triplet states descend from the same closed-shell reference.  Their
splitting therefore carries no bias from a spin-broken or unbalanced
determinant.  \pprpa{} has therefore demonstrated effective for diradicals, bond
dissociation and doubly excited states.
\cite{Yang2014Excitation,Li2023Linear,Yu2026LibppRPA}.

The principal obstacle to wider application is computational cost, and its
origin is the same as in TDDFT.  Both methods reduce to a generalized
eigenvalue problem over a space of orbital pairs, and in both the
two-electron integrals in the molecular-orbital basis, rather than the
diagonalization itself, dominate the operation count and the memory footprint.
What differs is which pairs span that space, and that is fixed by the reference
particle number. \cite{Tahir2019Comparing} Linear-response TDDFT conserves the electron number, so the
space is spanned by particle--hole pairs and has dimension
$N_{\mathrm{occ}}N_{\mathrm{vir}}$.  \pprpa{} changes the electron number by
two, so the space is spanned by particle--particle and hole--hole pairs and has
dimension $\mathcal{O}(N_{\mathrm{vir}}^{2})$ in the pp channel.  Since
$N_{\mathrm{vir}}\gg N_{\mathrm{occ}}$ in any converged basis, the \pprpa{}
pair space is the larger of the two, and a Davidson matrix--vector product
costs $\mathcal{O}(N_{\mathrm{aux}}N_{\mathrm{vir}}^{3})$ against
$\mathcal{O}(N_{\mathrm{aux}}N_{\mathrm{occ}}N_{\mathrm{vir}}^{2})$ for TDDFT.
The balanced treatment of both multiplicities is therefore bought at a cost
ratio of roughly $N_{\mathrm{vir}}/N_{\mathrm{occ}}$. 

Two families of approximation reduce this cost for excited states.  They differ
in the way to do the approximations.  The first replaces the two-electron integrals by
electrostatic interactions between atomic partial charges.  Niehaus \emph{et
al.} introduced this within tight binding \cite{Niehaus2001Tight}.  Grimme
\emph{et al.} carried it into Kohn--Sham TDDFT as the simplified Tamm--Dancoff
approach, in which L\"owdin monopoles interact through a damped atom--atom
kernel with the correct $1/R$ asymptote
\cite{Grimme2013Simplified,Bannwarth2014Simplified}.  R\"uger \emph{et al.}
built the same coupling from Kohn--Sham orbitals \cite{Ruger2016Tight}, and we
have previously replaced the two-centre integrals of the TDDFT coupling matrix
by a gamma function requiring no tight-binding parameters, together. \cite{Wang2024Implementation}.

The second family retains the resolution of the identity and reduces the
auxiliary basis.  Giannone \emph{et al.} \cite{Giannone2020Minimal} and Zhou
\emph{et al.} \cite{Zhou2023Minimal} represent the fitting set by a
\emph{minimal} number of atom-centred Gaussian shells, with exponents obtained
from tabulated atomic radii \cite{Ghosh2008Wave} scaled by a single global
factor; these are the TDDFT--as and TDDFT--ris methods.  The three-index tensor
retains the Coulomb metric, so the auxiliary-set size acts as a convergence
parameter, and adding shells recovers conventional density fitting.  Zhou
\emph{et al.} have combined the construction with graphics processing units to
reach organic molecules of several thousand atoms \cite{Zhou2026GPU}. 

The question this work addresses is whether large diradicals, molecular and
periodic alike, can be treated at a cost that makes them routine without giving
up accuracy.  We therefore interface \pprpa{} with the minimal auxiliary basis.
We then refer to this new method as \pprpa{}--AS, continuing the naming of the
TDDFT--as construction from which it derives \cite{Giannone2020Minimal}.  The
auxiliary set is specified by the shells it contains, $s$, $sp$ or $spd$;
unless stated otherwise the benchmark results below use $spd$.

We wanted to address one point here: the reference determinant comes from a conventional self-consistent field calculation and is used unmodified, so the minimal auxiliary basis enters only the \pprpa{} step.  Conventional and approximate results are therefore obtained on identical orbitals and differ by one approximation and nothing else. Compared ppRPA-AS with the TDDFT-as method: TDDFT expands particle--hole densities, while the
\pprpa{} pair space is built from virtual--virtual pairs. That difference makes the approximation worth
more here.  The pair space grows as $O(N_{\mathrm{vir}}^{2})$ rather than
$N_{\mathrm{occ}}N_{\mathrm{vir}}$, so more of the work carries the auxiliary
index. We calibrate the scaling factor for
\pprpa{}, measure how far it transfers, and benchmark on the 28 molecules of the
Thiel set, metal dimers, ligand-protected complexes of up to a hundred atoms,
and the 286-atom $\mathrm{Au_{25}}$ dimer.  We then apply the method under
periodic boundary conditions to seven point defects in four host materials,
among them the nitrogen--vacancy center in diamond \cite{Li2024Particle}.

\section{Theory and Methods}
\label{sec:theory_methods}

\subsection{Particle--Particle Random Phase Approximation}
\label{sec:pprpa_theory}

We summarize here the \pprpa{} equations required in what follows.  The
method itself is established, and its derivation is given elsewhere
\cite{VanAggelen2013Exchange,VanAggelen2014Exchange,Yang2013Benchmark,Yu2026LibppRPA}.

\begin{figure}[htbp]
  \centering
  \includegraphics[width=\linewidth]{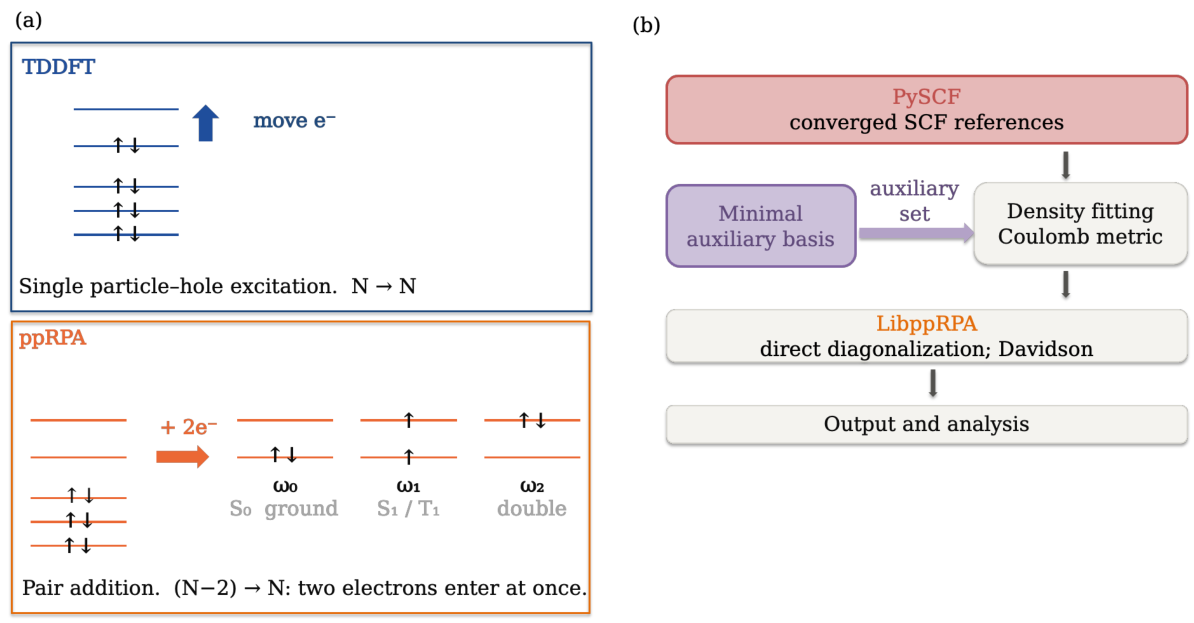}
  \caption{(a) Schematic representation of the difference between TDDFT and
    \pprpa{}--AS.  (b) Architecture of a \pprpa{}--AS calculation.}
  \label{fig:theory}
\end{figure}

\pprpa{} differs from linear-response TDDFT in its reference state
[Figure~\ref{fig:theory}(a)].  TDDFT is formulated about the $N$-electron
ground state, where $N$ is the number of electrons in the target system, and
propagates particle--hole amplitudes.  \pprpa{} responds
instead to a pair-creation or pair-annihilation field.  In the hole--hole (hh)
route used here, the reference is the closed-shell $(N+2)$-electron system, and
the neutral $N$-electron states follow by removing an electron pair.  The
particle--particle (pp) route starts from an $(N-2)$-electron reference and
adds a pair.

Let $i<j$ and $k<l$ index occupied and $a<b$ and $c<d$ virtual orbitals of the
charged reference, so that $\bm{A}$ has dimension
$N_{\mathrm{vir}}(N_{\mathrm{vir}}-1)/2$ and $\bm{C}$ has dimension
$N_{\mathrm{occ}}(N_{\mathrm{occ}}-1)/2$, with $N_{\mathrm{occ}}$ and
$N_{\mathrm{vir}}$ the numbers of occupied and virtual orbitals of that
reference.  In spin-orbital notation the
\pprpa{} equation is
\begin{equation}
 \begin{pmatrix}
  \bm{A} & \bm{B} \\
  \bm{B}^{\dagger} & \bm{C}
 \end{pmatrix}
 \begin{pmatrix}
  \bm{X}_{m} \\
  \bm{Y}_{m}
 \end{pmatrix}
 =
 \widetilde{\Omega}_{m}
 \begin{pmatrix}
  \bm{I} & \bm{0} \\
  \bm{0} & -\bm{I}
 \end{pmatrix}
 \begin{pmatrix}
  \bm{X}_{m} \\
  \bm{Y}_{m}
 \end{pmatrix}.
 \label{eq:pprpa_generalized}
\end{equation}
Here $\bm{A}$ and $\bm{C}$ are the particle--particle and hole--hole blocks,
$\bm{B}$ couples them, $\bm{I}$ and $\bm{0}$ are the identity and zero
matrices, and $m$ labels the \pprpa{} root.
The amplitudes $X_{ab}^{m}$ and $Y_{ij}^{m}$ weight creation of a pair in
virtual orbitals $a,b$ and removal of a pair from occupied orbitals $i,j$.
Pair-addition roots are dominated by $\bm{X}_{m}$, and the pair-removal roots
used here by $\bm{Y}_{m}$.

With $\widetilde{\epsilon}_{p}=\epsilon_{p}-\mu$, where $\epsilon_{p}$ is the
energy of orbital $p$ of the charged reference and $\mu$ is the chemical
potential, the required matrix blocks are
\begin{subequations}
\label{eq:pprpa_blocks}
\begin{align}
 A_{ab,cd}
 &=\delta_{ac}\delta_{bd}
   (\widetilde{\epsilon}_{a}+\widetilde{\epsilon}_{b})
   +\langle ab\Vert cd\rangle, \\
 B_{ab,kl}
 &=\langle ab\Vert kl\rangle, \\
 C_{ij,kl}
 &=-\delta_{ik}\delta_{jl}
   (\widetilde{\epsilon}_{i}+\widetilde{\epsilon}_{j})
   +\langle ij\Vert kl\rangle,
\end{align}
\end{subequations}
where $\delta_{ac}$ is the Kronecker delta and
$\langle pq\Vert rs\rangle=\langle pq|rs\rangle-\langle pq|sr\rangle$ is
an antisymmetrized two-electron integral in physicists' notation.  We take
$\mu=(\epsilon_{\mathrm{HOMO}}+\epsilon_{\mathrm{LUMO}})/2$, so that
$\widetilde{\Omega}_{m}=\Omega_m-2\mu$, where $\Omega_m$ is the two-electron
addition or removal energy of root $m$.  This shift is constant and cancels
from energy differences.  For a restricted closed-shell reference the equation
separates into singlet and triplet pair spaces \cite{Yang2013Benchmark}.

An individual \pprpa{} eigenvalue is a two-electron addition or removal energy
relative to the charged reference, not a neutral excitation energy.  For the
$(N+2)$-electron hh route,
\begin{equation}
 \begin{aligned}
  \Omega^{-}_{m}
  &=E_0^{N+2}-E_m^N, \\
  \Delta E_m^{\mathrm{hh}}
  &=E_m^N-E_0^N
    =\Omega^{-}_{0}-\Omega^{-}_{m},
 \end{aligned}
 \label{eq:hh_gap}
\end{equation}
where $E_m^N$ is the total energy of the $m$-th $N$-electron state and
$E_0^{N+2}$ that of the closed-shell reference,
so every target state is referenced to the $m=0$ root rather than to the total
energy of the charged determinant.  The $(N-2)$-electron pp route gives
\begin{equation}
 \begin{aligned}
  \Omega^{+}_{m}
  &=E_m^N-E_0^{N-2}, \\
  \Delta E_m^{\mathrm{pp}}
  &=\Omega^{+}_{m}-\Omega^{+}_{0}.
 \end{aligned}
 \label{eq:pp_gap}
\end{equation}
For a neutral target, the hh and pp references therefore carry charges $-2$
and $+2$.

\subsection{Auxiliary-$s$ Approximation}
\label{sec:auxiliary_s}

The auxiliary-$s$ (AS) approximation replaces the auxiliary basis used for the
two-electron integrals by a minimal atom-centered set.  In its basic form one
uncontracted $s$-type Gaussian is placed on every atom.  The function centered
at $\bm{R}_A$, the position of atom $A$, is
\begin{equation}
 \chi_A(\bm{r})
 =N_A(\alpha_A)
  \exp\!\left[-\alpha_A
  \lvert\bm{r}-\bm{R}_A\rvert^2\right],
 \label{eq:as_function}
\end{equation}
where $N_A$ is the normalization constant and $\bm{r}$ the electronic
coordinate.  Its exponent follows from a
free-atom radius,
\begin{equation}
 \alpha_A(\theta)
 =\theta\,\xi_A
 =\frac{\theta}{R_A^2},
 \label{eq:as_exponent}
\end{equation}
with $R_A$ the scalar free-atom radius of element $A$ in bohr, not to be
confused with the position $\bm{R}_A$ of Eq.~\eqref{eq:as_function},
$\xi_A=R_A^{-2}$ tabulated per element, and $\theta$ a
global dimensionless scaling parameter
\cite{Ghosh2008Wave,Giannone2020Minimal,Zhou2023Minimal}.  Atoms of the same
element therefore share one exponent at a given $\theta$, while different
elements retain different radial extents.

The single $s$ function is the smallest member of a nested family.  The $sp$
variant adds one $p$ shell to every non-hydrogen atom, and $spd$ adds a $d$
shell as well, both using the exponent of Eq.~\eqref{eq:as_exponent}.  With
spherical Gaussians, and hydrogen carrying only the $s$ function,
\begin{equation}
  N_{\mathrm{aux}} = n_{\mathrm{sh}} N_{\mathrm{heavy}} + N_{\mathrm{H}},
  \qquad n_{\mathrm{sh}} = 1,\, 4,\, 9
  \ \text{for } s,\ sp \text{ and } spd.
  \label{eq:naux}
\end{equation}
Here $N_{\mathrm{heavy}}$ and $N_{\mathrm{H}}$ are the numbers of non-hydrogen
and hydrogen atoms and $n_{\mathrm{sh}}$ the number of auxiliary functions per
heavy atom.
$N_{\mathrm{aux}}$ is therefore fixed by atomic composition alone and is
decoupled from the orbital basis size, which is the origin of the reductions
reported below.

The auxiliary functions enter through the usual Coulomb metric.  For orbital
basis functions $\phi_\mu$ and auxiliary functions $\chi_P$,
\begin{equation}
 J^{P}_{\mu\nu}=(\mu\nu|P),
 \qquad
 V_{PQ}=(P|Q).
 \label{eq:as_three_center_metric}
\end{equation}
Here $P$ and $Q$ label auxiliary functions and $\mu$ and $\nu$ orbital basis
functions.
Metric orthonormalization gives
\begin{equation}
 B^{P}_{\mu\nu}
 =\sum_Q (\bm{V}^{-1/2})_{PQ}J^{Q}_{\mu\nu},
 \label{eq:as_ao_tensor}
\end{equation}
and the four-index integrals in the \pprpa{} blocks are approximated by
\begin{equation}
 L^{P}_{pq}
 =\sum_{\mu\nu}C_{\mu p}C_{\nu q}B^{P}_{\mu\nu}.
 \label{eq:as_lpq}
\end{equation}
The coefficients $C_{\mu p}$ are the molecular orbitals of the charged
reference expanded in the orbital basis, and $p$ and $q$ run over all of them.
These integrals, and their exchange permutations, enter the $\bm{A}$,
$\bm{B}$ and $\bm{C}$ blocks of Eq.~\eqref{eq:pprpa_generalized}.

Nothing else is altered.  The charged-reference SCF solution, the molecular
orbitals and their energies, the chemical potential, the pair space, the spin
adaptation and the eigensolver are all those of the conventional \pprpa{}
calculation.  The AS error can therefore be isolated on identical orbitals and
numerical settings,
\begin{equation}
 \delta_m(\theta)
 =\Delta E_m^{\mathrm{AS}}(\theta)-\Delta E_m^{\mathrm{conv}},
 \label{eq:as_error}
\end{equation}
with each excitation energy referenced to the ground-state \pprpa{} root
obtained from the same auxiliary set.

We have extended the auxiliary-$s$ approximation from TDDFT to \pprpa{}.  The
implementation is in PySCF, with the \pprpa{} eigenvalue problem solved through
LibppRPA \cite{Sun2018PySCF,Sun2020Recent,Yu2026LibppRPA,Sun2026PySCF}
[Figure~\ref{fig:theory}(b)].

\section{Benchmarking}
\label{sec:Benchmarking}

The method is assessed on 32 molecules, from a diatomic to a 286-atom
nanocluster, and on seven point defects in four crystalline hosts.  The 28
molecules of the Thiel set \cite{Schreiber2008Benchmarks} fix the behavior for
organic valence excitations.  They are run with two auxiliary sets, $asp$ and
$aspd$, which separates the effect of angular flexibility in the auxiliary basis
from the effect of system size.  Three metal-containing molecules follow: the
gold dimer \cite{Bishea1991Spectroscopic}, a 100-atom phosphine-bridged digold
cation \cite{Cao2006Bis}, and an 84-atom zinc bis(semiquinone) complex
\cite{Piskunov2011Zinc}.  Minimal auxiliary bases of this type were constructed
and optimized on organic molecules \cite{Zhou2023Minimal,Zhou2026GPU}, and the
metal-containing case has been addressed only for silver nanoparticles with a
single $s$ function per atom \cite{Giannone2020Minimal}.  The 286-atom
$\mathrm{Au_{25}}$ dimer \cite{Swierczewski2023Exceptionally} then places the
method at the size it is intended for, and seven point defects in diamond,
hexagonal BN, MgO, and SiC \cite{Li2024Particle} extend it to periodic boundary
conditions.  Every system is computed twice from the same converged reference
orbitals and orbital energies, once with a conventional \pprpa{} set and
once with the minimal auxiliary basis.  Accuracy is reported as the error in the
within-method excitation gap for the lowest singlet and triplet roots, and
$\theta$ is scanned for each system and each reference determinant.

To set the scan range we start from a target width. We want the scaling parameter to put the auxiliary functions in a physically meaningful range. $\theta$ fixes the width of the auxiliary function on each atom, $d_A = R_A/\sqrt{\theta}$. Each auxiliary function represents a density, and a density is a product of two orbitals, so its exponent is the sum of the two orbital exponents. For carbon in def2-TZVP these sums have a median of $0.91$~bohr$^{-2}$ and an interquartile range of $0.55$ to $1.33$. Matching the median places the $spd$ guess near $1.4$. All shells share one exponent but peak at different radii. The radial density of an $l$-type primitive peaks at $\sqrt{(l+1)/2\alpha}$, so moving from $s$ to $spd$ carries the mean peak radius outward by about $1.4$. The exponent rises by the square of that, which puts the $s$ guess near $0.7$. We scan a few points spanning $\pm 30\%$ about the guess. Meanwhile, the metal-containing systems are different. There the metal--metal and metal--ligand interactions can move the optimum, so we scan a broader range and discuss the choice of $\theta$ for such systems in detail in Section~\ref{sec:guidelines}.

For the SCF calculations, the reference determinants are
obtained with Hartree--Fock and with the BP86,\cite{Becke1988Density,%
Perdew1986Density} PBE,\cite{Perdew1996Generalized}
B3LYP\cite{Becke1993Density,Lee1988Development} and
PBE0\cite{Adamo1999Toward} functionals: HF, B3LYP and PBE0 for the Thiel set;
HF, BP86, B3LYP and PBE0 for the metal complexes; PBE and PBE0 for the
$\mathrm{Au_{25}}$ dimer; and PBE and B3LYP for the point defects.  The Thiel
set and the metal-containing molecules use the def2-TZVP orbital
basis,\cite{Weigend2005Balanced} with the associated effective core potentials
on gold;\cite{Andrae1990Energy} copper is treated all-electron, as
the def2 family defines no copper potential.  The $\mathrm{Au_{25}}$ dimer uses
def2-SVP with the corresponding gold potential, and the point defects use
cc-pVDZ.\cite{Dunning1989Gaussian}  The conventional \pprpa{} calculation
against which the approximation is measured fits the two-electron integrals
with def2-TZVP-JKFIT\cite{Weigend2008Hartree} for the molecular systems, with
the def2-universal Coulomb-fitting set\cite{Weigend2006Accurate} for the
$\mathrm{Au_{25}}$ dimer, and with cc-pVDZ-RI\cite{Weigend2002Efficient} for
the periodic systems; the reference self-consistent field calculation is
density fitted with the def2-universal set throughout.  The minimal auxiliary
basis replaces the fitting set of the \pprpa{} step alone, so each comparison
is made within one fitting convention.  The \pprpa{} eigenvalue problem is
solved with LibppRPA.\cite{Yu2026LibppRPA}

\section{Results and Discussion}
\label{sec:results}

\subsection{Benchmark Results}
\label{sec:benchmarkresults}

\subsubsection{Thiel Tests}
\label{sec:thiel_Tests}

\begin{figure}[htbp]
  \centering
  \includegraphics[width=\linewidth]{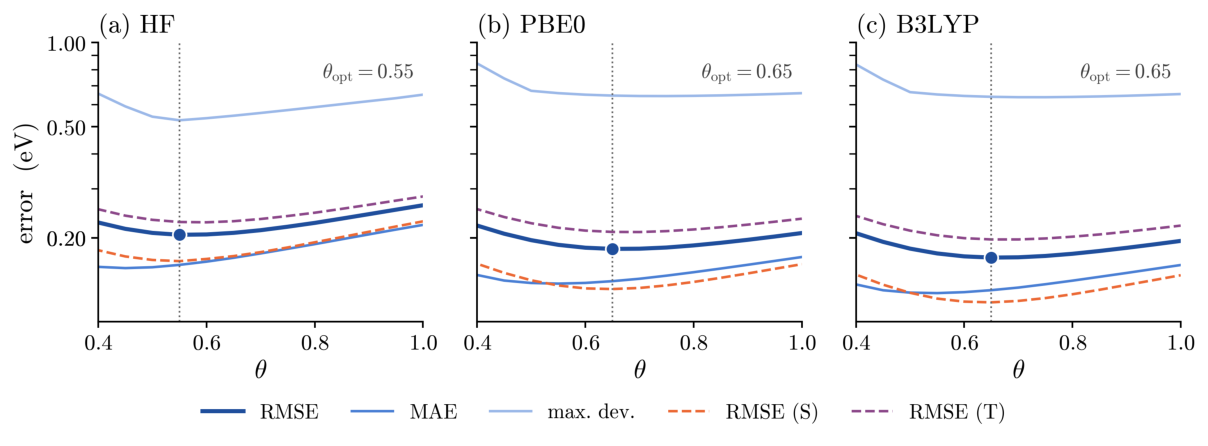}
  \caption{Error of \pprpa{}--AS relative to conventional \pprpa{} on the Thiel
    set with the $sp$ auxiliary set, as a function of the scaling parameter
    $\theta$.  Panels are (a) HF, (b) PBE0 and (c) B3LYP.  Each shows the RMSE,
    the MAE, the largest single deviation, and the RMSE resolved into singlet
    and triplet states.  Every curve covers 140 states per reference: 28
    molecules, five states each ($S_1$, $S_2$ and $T_1$--$T_3$).  All values are
    in eV on a logarithmic scale.  The dotted line marks
    $\theta_{\mathrm{opt}}$ for that reference.}
  \label{fig:thiel}
\end{figure}

The Thiel set is a standard benchmark for excited-state methods and comprises
28 molecules built from light elements, so no effective core potentials are
required.  Table~\ref{tab:thiel28} compares \pprpa{}--AS with conventional
\pprpa{} on this set.  Both calculations start from the same converged reference
orbitals and orbital energies, so the auxiliary basis is the only difference
between them.  We start with the $sp$ set.  At the shared optimum
$\theta = 0.65$ the root-mean-square error is $0.187$~eV over 420 excitation
energies: 28 molecules, five states each ($S_1$, $S_2$ and $T_1$--$T_3$), for
each of the three reference determinants (Figure~\ref{fig:thiel}).

The accuracy at the optimum barely depends on which reference is used.  The
root-mean-square error is $0.205$~eV for Hartree--Fock, $0.170$~eV for B3LYP and
$0.182$~eV for PBE0, a spread of $0.035$~eV between the largest and the
smallest.  The minimum is shallow: $\theta = 0.60$ gives $0.1876$~eV against
$0.1875$~eV at $0.65$.  The position of the optimum does depend on the
reference.  Hartree--Fock optimizes at $\theta = 0.55$ and both hybrids at
$0.65$, so $\theta_{\mathrm{opt}}$ rises as the fraction of exact exchange
decreases.

\begin{table}[htbp]
  \centering
  \caption{Accuracy of \pprpa{}--AS against conventional \pprpa{} on the
    Thiel benchmark set.  Both calculations use the same converged reference
    orbitals and orbital energies.  Errors are as minus conventional \pprpa{}
    excitation energies over the 28 molecules, five states each (S$_1$, S$_2$,
    T$_1$--T$_3$); 140 states per reference.  All entries in eV;
    $\theta_{\mathrm{opt}}$ minimizes the RMSE for that reference, and the last
    row is evaluated at the shared optimum over all 420 states.}
  \label{tab:thiel28}
  \begin{tabular}{lcccc}
    \toprule
    Reference & $\theta_{\mathrm{opt}}$ & MAE & RMSE & Max \\
    \midrule
    HF        & 0.55 & 0.160 & 0.205 & 0.527 \\
    B3LYP     & 0.65 & 0.130 & 0.170 & 0.639 \\
    PBE0      & 0.65 & 0.140 & 0.182 & 0.646 \\
    \midrule
    All three & 0.65 & 0.146 & 0.187 & 0.646 \\
    \bottomrule
  \end{tabular}
\end{table}

The largest deviations are concentrated in a single chemical family.
Formaldehyde, acetone, acetamide and propanamide head the RMSE ranking at
$0.461$, $0.337$, $0.327$ and $0.293$~eV, an order of magnitude above
naphthalene at $0.043$~eV (Table~\ref{tab:thiel28_outliers}).  Two effects
produce this concentration.  First, the auxiliary set is fixed by atomic
composition through Eq.~\eqref{eq:naux}, not by the orbital basis and not by the
complexity of the density it has to represent.  Formaldehyde therefore carries
10 auxiliary functions and naphthalene 48, although the pair densities in the
two molecules are of comparable radial complexity.  A small molecule also has no
redundancy: with few centers, a poorly chosen exponent on one atom cannot be
compensated by its neighbors, as it can in a larger system.  The molecular RMSE
correlates with $N_{\mathrm{aux}}$ with a coefficient of $-0.68$ across the set.
Second, the carbonyls remain substantially worse than the $\pi$ systems even at
matched auxiliary size, $0.336$ against $0.158$~eV for
$N_{\mathrm{aux}} \le 30$.  The compact $n \rightarrow \pi^{*}$ transition
density is represented poorly by a minimal atom-centered expansion.

Two further biases appear, both specific to the $sp$ set.  Triplet states are
less accurate than singlets: the triplet RMSE exceeds the singlet RMSE by a
factor of $1.3$ to $1.7$ for every reference and at every $\theta$ examined,
$0.197$ against $0.117$~eV for B3LYP at its optimum.  All five of the largest
single deviations are triplets.  Four of the five come from the Hartree--Fock
reference, the least accurate of the three at its own optimum, $0.205$~eV
against $0.170$ and $0.182$~eV.

\begin{table}[htbp]
  \centering
  \caption{Molecules with the largest and smallest as deviations on the
    Thiel set at the shared $\theta=0.65$, averaged over the HF, B3LYP and
    PBE0 references with 15 states each.  $N_{\mathrm{aux}}$ is the auxiliary
    basis size with the $sp$ shell set.  Energies in eV.}
  \label{tab:thiel28_outliers}
  \begin{tabular}{lcccc}
    \toprule
    Molecule & $N_{\mathrm{aux}}$ & RMSE & Max dev. & Mean signed \\
    \midrule
    \multicolumn{5}{l}{\textit{Largest deviations}}\\
    Formaldehyde    & 10 & 0.461 & $+0.646$ & $+0.436$ \\
    Acetone         & 22 & 0.337 & $+0.509$ & $+0.325$ \\
    Acetamide       & 21 & 0.327 & $+0.503$ & $+0.308$ \\
    Propanamide     & 27 & 0.293 & $+0.471$ & $+0.282$ \\
    Ethene          & 12 & 0.219 & $+0.308$ & $+0.170$ \\
    \midrule
    \multicolumn{5}{l}{\textit{Smallest deviations}}\\
    Thymine         & 42 & 0.065 & $+0.107$ & $+0.039$ \\
    E-hexatriene    & 32 & 0.059 & $+0.085$ & $+0.031$ \\
    E-octatetraene  & 42 & 0.047 & $+0.065$ & $+0.028$ \\
    Naphthalene     & 48 & 0.043 & $+0.059$ & $+0.010$ \\
    \bottomrule
  \end{tabular}
\end{table}

\begin{figure}[htbp]
  \centering
  \includegraphics[width=\linewidth]{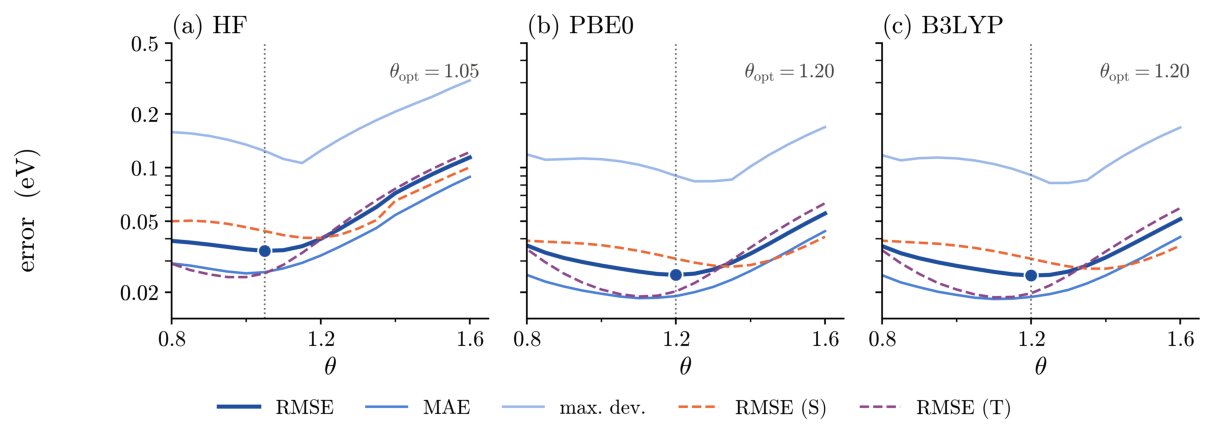}
  \caption{Error of \pprpa{}--AS relative to conventional \pprpa{} on the Thiel
    set with the $spd$ auxiliary set, as a function of the scaling parameter
    $\theta$.  Panels are (a) HF, (b) PBE0 and (c) B3LYP.  Each shows the RMSE,
    the MAE, the largest single deviation, and the RMSE resolved into singlet
    and triplet states, in eV on a logarithmic scale.  The dotted line marks
    $\theta_{\mathrm{opt}}$: $1.05$ for Hartree--Fock and $1.20$ for both
    hybrids.  Singlets are less accurate than triplets near the optimum.  This
    reverses the ordering obtained with $sp$ (Figure~\ref{fig:thiel}). }
  \label{fig:thiel_spd}
\end{figure}

The rest of this work uses the $spd$ set.  Repeating the benchmark with $spd$,
on the same geometries, orbitals and active spaces, reduces the error by a
factor of $6.4$: the shared optimum moves to $\theta = 1.10$ and the
root-mean-square error over the same 420 excitation energies falls from $0.187$
to $0.029$~eV.  The individual references give $0.034$~eV at $\theta = 1.05$ for
Hartree--Fock, and $0.025$~eV at $\theta = 1.20$ for both B3LYP and PBE0 (Figure~\ref{fig:thiel_spd}).

The $d$ functions also change which molecules fail, and all three biases of the
$sp$ set either weaken or reverse.  Under $sp$ the carbonyls dominate the error
ranking, which we attributed to the compact $n \rightarrow \pi^{*}$ transition
density that an atom-centered $s$ and $p$ expansion represents poorly.
Supplying the missing angular flexibility removes the effect: formaldehyde falls
from first place at $0.461$~eV to sixth at $0.034$~eV, and the carbonyl excess
over $\pi$ systems at matched auxiliary size disappears.  The singlet--triplet
ordering reverses: singlets are now less accurate than triplets, by a factor of
$1.6$ at the shared optimum, and the largest single deviations are the ethene
$S_1$ energies.  The systematic overestimation also disappears, from 27 of 28
molecules with a positive mean signed error under $sp$ to 12 of 28 under $spd$.
What remains is a size effect.  The correlation
between molecular RMSE and $N_{\mathrm{aux}}$ strengthens from $-0.68$ to
$-0.77$, and the largest deviations now belong to the smallest molecules:
cyclopropene at $0.058$~eV with 31 auxiliary functions and ethene at $0.056$~eV
with 22, against adenine at $0.013$~eV with 95 and naphthalene at $0.016$~eV
with 98.

\subsubsection{Metal Complexes}
\label{sec:metal_complexes}

The Thiel set examined in Section~\ref{sec:thiel_Tests} comprises light
main-group elements only.  Metal-containing systems differ from it in three
ways that bear on the as approximation.  Their valence orbitals are more
diffuse, so the pair densities the auxiliary set must represent are more
extended.  The heavier elements require effective core potentials, which change
the radial structure of the orbitals being fitted.  And the excitations of
interest may be metal-centered or ligand-centered, so the element that has to
be described well is not fixed by the formula of the complex.

The exponents compound these differences.  Each is set by a free-atom radius
through Eq.~\eqref{eq:as_exponent}, and for a heavy element that radius does not
reflect how diffuse the valence orbitals actually are.  Relativistic contraction
makes the gold atom compact, $R_{\mathrm{Au}} = 0.524$~\AA{} against
$R_{\mathrm{C}} = 0.651$~\AA{}, so the table assigns gold the tighter auxiliary
function even though its $5d$ and $6s$ orbitals are the more diffuse of the two.
One global scale factor acts on every element at once, so a value optimal for
organic molecules need not transfer.  We therefore examine three
metal-containing systems, chosen to vary two properties independently: which
element carries the excitation, and how large an auxiliary set the molecule
supports.

\begin{figure}[htbp]
  \centering
  \includegraphics[width=\linewidth]{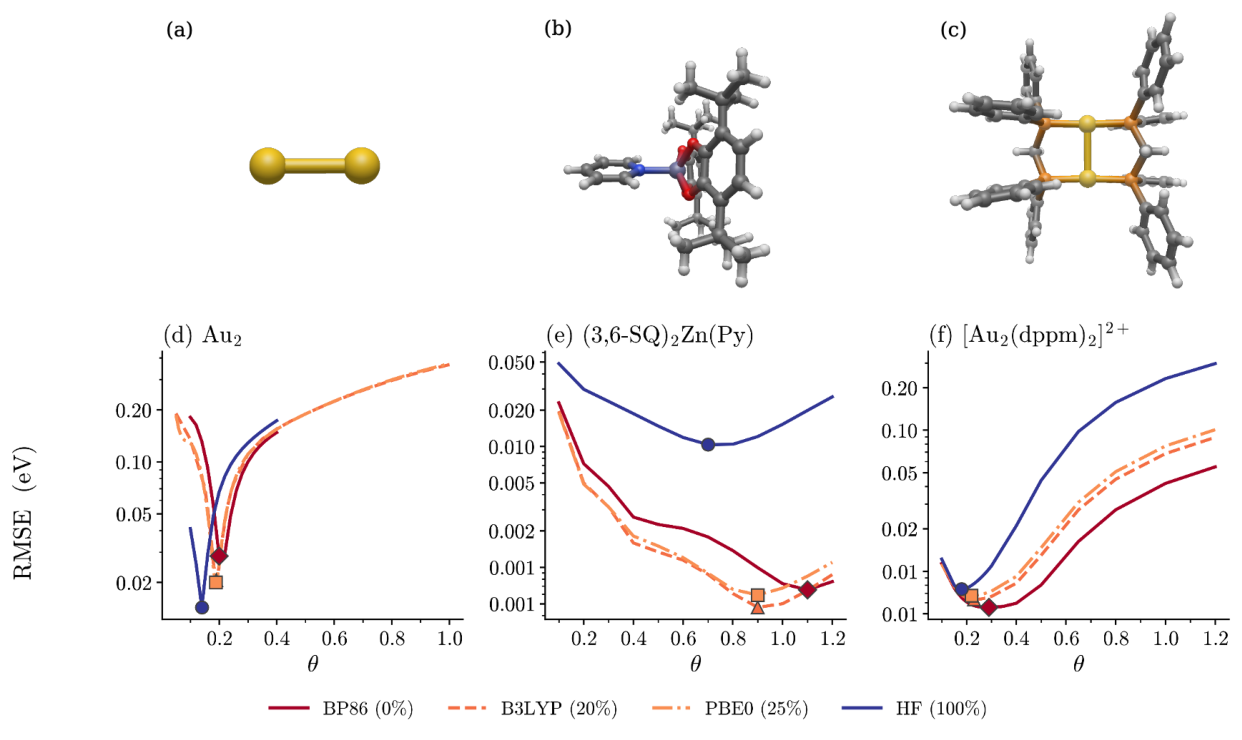}
  \caption{The three metal-containing systems and the dependence of the as
    error on the scaling parameter.  Each column corresponds to one system.
    (a)--(c) Structures of $\mathrm{Au_2}$,
    $\mathrm{(3,6\text{-}SQ)_2Zn(Py)}$ and
    $\mathrm{[Au_2(dppm)_2]^{2+}}$, comprising 2, 84 and 100 atoms and
    carrying 18, 396 and 548 auxiliary functions respectively.
    (d)--(f) Error of \pprpa{}--AS relative to conventional \pprpa{} for the
    same three systems as a function of $\theta$.  Curves are colored by
    exact-exchange fraction, from BP86 at 0\% to Hartree--Fock at 100\%, and
    filled markers indicate the minimum for each reference.  Axis ranges
    differ between panels: both $\theta_{\mathrm{opt}}$ and the error scale
    vary substantially across the three systems.  Numerical values are given
    in Table~S1.}
  \label{fig:metal_complexes}
\end{figure}

\begin{itemize}
\item \textbf{$\mathrm{Au_2}$.}\cite{Bishea1991Spectroscopic}  A diatomic with
      no ligand shell, so the excitation is entirely metal-centered.  Its
      auxiliary set of 18 functions is the smallest considered here.  It is the
      limiting case of minimal auxiliary coverage, with no surrounding ligand
      density available to compensate a poorly chosen exponent.

\item \textbf{$\mathrm{(3,6\text{-}SQ)_2Zn(Py)}$.}\cite{Piskunov2011Zinc}
      Zinc(II) is a closed-shell $d^{10}$ center and the two unpaired electrons
      reside on the semiquinone ligands, so the excitation is ligand-centered
      although the system is a metal complex.  With 396 auxiliary functions, it
      separates the presence of a metal from whether the metal carries the
      states of interest.

\item \textbf{$\mathrm{[Au_2(dppm)_2]^{2+}}$.}\cite{Cao2006Bis}  Metal-centered
      as in $\mathrm{Au_2}$, with low-lying excitations arising from the
      aurophilic $\mathrm{Au}-\mathrm{Au}$ interaction.  It is
      ligand-protected, comprises 100 atoms, and carries 548 auxiliary
      functions.  It therefore separates which element carries the excitation
      from the size of the auxiliary set.
\end{itemize}

All three metal complexes use the $spd$ auxiliary basis set, the results is shown in Figure~\ref{fig:metal_complexes}.

Over the four references the as approximation reproduces conventional \pprpa{}
excitation energies to $0.0056$--$0.0075$~eV for the digold cation and to
$0.014$--$0.028$~eV for $\mathrm{Au_2}$.  The zinc complex separates into two
groups: $0.00047$--$0.00065$~eV for the three density functionals, and
$0.0104$~eV for Hartree--Fock, a factor of 16 worse and the largest spread of
any system here.  The auxiliary reductions are $89.9\%$ for the zinc complex
and $90.1\%$ for the digold cation.  $\mathrm{Au_2}$, the smallest system
examined, is the least accurately reproduced under every reference.

The three systems also differ in how sharply the error depends on $\theta$.
$\mathrm{Au_2}$ has a sharp minimum, while the zinc complex and the digold
cation have shallow ones, so the two larger systems tolerate a much less careful
choice of $\theta$ than the diatomic does.

$\theta_{\mathrm{opt}}$ increases monotonically as exact exchange is removed,
in all three systems: $0.14$, $0.19$, $0.19$ and $0.20$ for $\mathrm{Au_2}$
under Hartree--Fock, PBE0, B3LYP and BP86; $0.70$, $0.90$, $0.90$ and $1.10$
for the zinc complex; and $0.18$, $0.22$, $0.23$ and $0.29$ for the digold
cation.  The same ordering appears in the Thiel set, where Hartree--Fock
optimizes at $\theta = 1.05$ against $1.20$ for both hybrids.

The \pprpa{} equations use bare Coulomb integrals, so the functional cannot act
through an exchange--correlation kernel.  It acts only by changing the
orbitals.  The particle--particle channel fits virtual--virtual pair densities,
and Hartree--Fock virtual orbitals are more diffuse than Kohn--Sham ones: a
Hartree--Fock virtual moves in the field of all $N$ electrons, whereas a
Kohn--Sham virtual sees the same local potential as the occupied orbitals and
remains valence-like.  A more diffuse pair density requires a broader auxiliary
Gaussian, and a broader Gaussian is one of smaller exponent, hence of smaller
$\theta$.  We confirmed this on $\mathrm{Au_2}$: the virtual--virtual pair
densities are $15\%$ more extended under Hartree--Fock than under BP86, their
extent tracks $\theta_{\mathrm{opt}}^{-1/2}$ to within $5\%$ across all four
functionals, and the occupied orbitals change by only $0.2\%$.

For the zinc complex the error is independent of the size of the gap being
reproduced.  The singlet--triplet separation varies from $3.26$ to $3.95$~eV
across the four references, while the as error stays below $0.001$~eV for the
three density functionals, at $\theta_{\mathrm{opt}} = 0.90$ for PBE0 and B3LYP
and $1.10$ for BP86.

The tolerance in $\theta$ increases with the size of the auxiliary set.  We
define it as the range of $\theta$ over which the RMSE remains within $10\%$ of
its minimum, expressed as a fraction of $\theta_{\mathrm{opt}}$.  It spans
$3$--$12\%$ for $\mathrm{Au_2}$, which carries 18 auxiliary functions,
$13$--$33\%$ for the zinc complex with 396, and $47$--$73\%$ for the digold
cation with 548.  Overlap among the auxiliary Gaussians produces this: where
many centers are present, a poorly chosen exponent on one center is partly
compensated by its neighbors, and a two-atom system affords no such redundancy.
$\mathrm{Au_2}$ is accordingly both the least accurately reproduced system in
Table~S1 and the most sensitive to $\theta$.  The approximation therefore
becomes more robust as the system grows.

\subsubsection{Large Diradical Metal Nanoclusters}
\label{sec:au25}

\begin{table}[htbp]
  \centering
  \caption{Lowest singlet--triplet separation of the covalently linked
    $\mathrm{Au_{25}}$ dimer, compared across methods.  The \pprpa{}
    calculations use the hole--hole channel from the closed-shell dianion, the
    def2-SVP orbital basis and a 40/40 active space; $N_{\mathrm{aux}}$ is the
    number of auxiliary functions.}
  \label{tab:au25}
  \begin{tabular}{llcc}
    \toprule
    Method & Reference & $N_{\mathrm{aux}}$ &
      $\Delta E_{\mathrm{ST}}$ (cm$^{-1}$) \\
    \midrule
    Broken-symmetry DFT\textsuperscript{\emph{a}}
                                  & PBE & ---   & $-604$ \\
    \pprpa{}                      & PBE    & 11450 & 0.0132\textsuperscript{\emph{b}} \\
    \pprpa{}--AS, $\theta = 0.20$ & PBE    & 1598  & 0.0120\textsuperscript{\emph{b}} \\
    \pprpa{}                      & PBE0   & 11450 & 0.0030\textsuperscript{\emph{b}} \\
    \pprpa{}--AS, $\theta = 0.20$ & PBE0   & 1598  & 0.0026\textsuperscript{\emph{b}} \\
    \bottomrule
  \end{tabular}

  \vspace{2pt}
  \raggedright\footnotesize
  \textsuperscript{\emph{a}} Data from Ref.~\citenum{Swierczewski2023Exceptionally};
  the open-shell singlet obtained by broken-symmetry LS-ROKS.\\
  \textsuperscript{\emph{b}} Of the order of the SCF convergence threshold
  ($10^{-8}$~Hartree, or $0.0022~\mathrm{cm^{-1}}$), and therefore an upper
  bound rather than a measured separation.
\end{table}

The largest system we have examined in this work is a covalently linked
$\mathrm{Au_{25}}$ dimer of 286 atoms, the
$\mathrm{[Au_{25}(SMe)_{16}]}$--diBINAS--$\mathrm{[Au_{25}(SMe)_{16}]}$ model
of the dimers reported by Swierczewski \emph{et al.}
\cite{Swierczewski2023Exceptionally}, reached through the hole--hole channel
from its closed-shell dianion.

We adopt $\theta = 0.20$, which lies in the narrow interval, $0.18$–$0.20$, common to the optima of both gold systems.

The AS approximation uses 1598 auxiliary functions against 11450 for the
conventional \pprpa{}, a reduction of $86\%$.  At $\theta = 0.20$ it reproduces
the conventional density-fitted excitation energies to a root-mean-square
deviation of $0.005~\mathrm{cm^{-1}}$ under PBE and $0.003~\mathrm{cm^{-1}}$
under PBE0, the closest agreement obtained for any system in this work. 

Both treatments place the lowest singlet and triplet below the resolution of
the calculation.  The separations in Table~\ref{tab:au25} are of the order of
the $0.0022~\mathrm{cm^{-1}}$ corresponding to the SCF convergence threshold,
and we therefore read them as upper bounds and assign no ordering to the two
multiplicities.
The near-degeneracy survives enlargement of the active space from 40 to 80
occupied and virtual orbitals, over which the separation moves only from 0.0126
to $0.0159~\mathrm{cm^{-1}}$.

A broken-symmetry treatment of the same model gives a very different answer.
DFT-PBE places the triplet $604~\mathrm{cm^{-1}}$ below the open-shell singlet
\cite{Swierczewski2023Exceptionally}, at least four orders of magnitude above
the present result.

\begin{figure}[htbp]
  \centering
  \includegraphics[width=\linewidth]{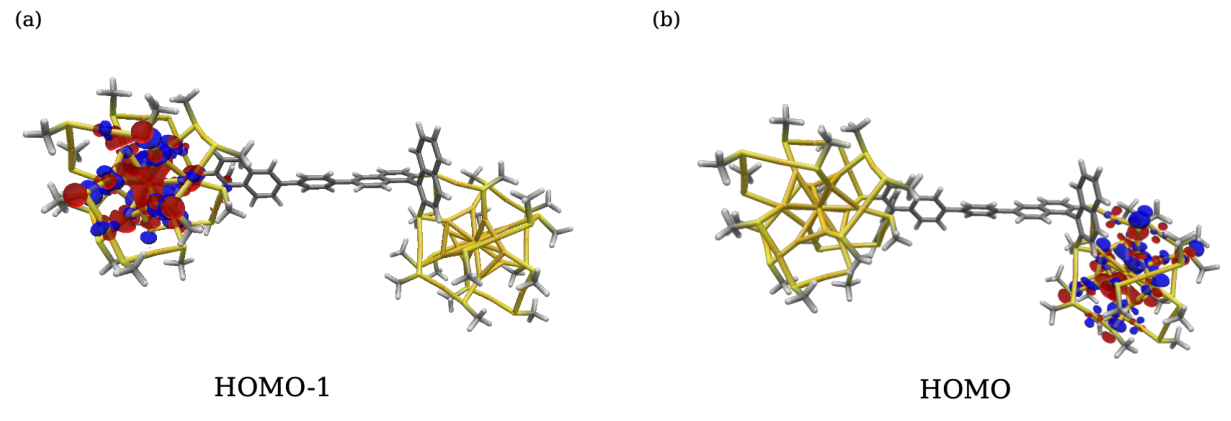}
  \caption{Frontier orbitals of the $\mathrm{Au_{25}}$ dimer: (a) HOMO$-1$ and
    (b) HOMO, from the PBE reference, plotted at an isosurface value of
    0.02~a.u.  Each orbital carries better than $99.9\%$ of its density on one
    of the two clusters. Details of densities localizations are shown in Table~S2}
  \label{fig:au25}
\end{figure}

Three observations indicate that the two unpaired electrons are magnetically
independent rather than coupled by the $604~\mathrm{cm^{-1}}$ obtained from the
broken-symmetry treatment.  First, the two orbitals do not overlap.
Each carries better than $99.9\%$ of its density on a single cluster and the
cluster centers are 25.5~\AA{} apart, so there is no pathway by which an
electron on one cluster can reach the other (Figure~\ref{fig:au25}).

Second, the \pprpa{} eigenvectors show the consequence directly.  The lowest
singlet is $99.9\%$ a single configuration in which one electron is removed
from each of the two frontier orbitals, and the configurations in which both
electrons are removed from the same orbital -- leaving both holes on one
cluster -- contribute $0.1\%$ or less.  It is this absence, rather than the
degeneracy itself, that carries the information: the triplet is restricted by
spin symmetry to configurations built from two distinct orbitals, so only the
singlet was free to admit the same-orbital configurations, and it does not.
Exchange coupling between two centers arises from precisely that admixture, and
it is absent here.

Third, broken-symmetry density functional theory overestimates exchange
couplings through spin contamination, whereas \pprpa{} obtains both
multiplicities from a single closed-shell reference and requires neither
symmetry breaking nor spin projection.  Consistent with all three, the reported
EPR and HYSCORE spectra of the monomer and the dimer are near-identical, from
which the same authors infer weak-to-vanishing intercluster coupling
\cite{Swierczewski2023Exceptionally}.

These results establish that \pprpa{} can treat diradicals at the scale
encountered in molecular quantum information science: for the 286-atom
$\mathrm{Au_{25}}$ dimer it describes the open-shell singlet and the triplet on
an equal footing from a single closed-shell reference, without symmetry
breaking or spin projection.The as approximation reproduces that description -- excitation energies and
configurational composition alike -- while discarding $86\%$ of the auxiliary
functions, which is what brings systems of this size within reach.

\subsubsection{Point Defects in Solids}
\label{sec:periodic}

The systems considered so far are molecular.  Point defects in semiconductors
and insulators pose the same physical problem -- a small number of electrons in
near-degenerate, spatially localized states -- in a periodic setting, and they
are the systems for which optically addressable spin qubits are actually
built.\cite{Li2024Particle}  We therefore apply \pprpa{}--AS to seven defects
in four host materials, using the geometries of
Ref.~\citenum{Li2024Particle}.

Three are in diamond: the nitrogen--vacancy center in its $^{3}A_{2}$ geometry,
the silicon--vacancy center, and the neutral carbon vacancy in two distinct
geometry branches, all in 64-atom supercells.  The remainder are the
$\mathrm{C_B C_N}$ pair in a 72-atom hexagonal boron nitride cell, the oxygen
vacancy in a 54-atom MgO cell, and the divacancy in a 126-atom SiC cell.
Charged closed-shell references are used throughout, through the hole--hole
channel for the nitrogen--vacancy, silicon--vacancy and divacancy centers and
the particle--particle channel for the others.  Both PBE and B3LYP references
are treated.  The active space begins at 200 occupied and 200 virtual orbitals
and is expanded at either boundary so that no manifold split by less than
$10^{-4}$~Hartree is divided.  Conventional and approximate calculations use
the same converged orbitals, and roots are matched one-to-one by maximum
eigenvector overlap within each spin manifold.

Table~S3 reports the error of \pprpa{}--AS against
conventional \pprpa{} at the shared optimum $\theta = 1.15$, over both
references.  The root-mean-square deviation across 58 excitation energies is
$0.0054$~eV and the largest single deviation is $0.020$~eV.  The accuracy is
therefore comparable to that obtained for the molecular metal complexes of
Section~\ref{sec:metal_complexes}, in supercells of 54 to 126 atoms and under
periodic boundary conditions.  Six of the seven defects fall below
$0.0037$~eV; the nitrogen--vacancy center is in a band of its own at
$0.0116$~eV, and it is the defect the quantum-information literature cares
about most.

\begin{figure}[htbp]
  \centering
  \includegraphics[width=\linewidth]{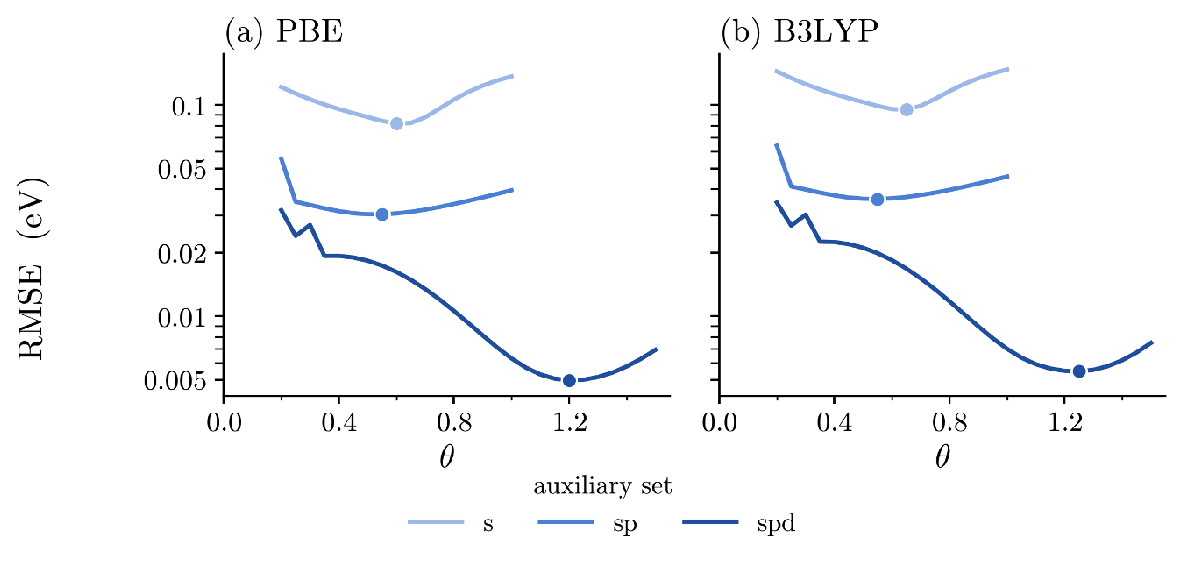}
  \caption{Error of \pprpa{}--AS relative to conventional \pprpa{} for seven
    point defects, as a function of the scaling parameter $\theta$, for the
    (a) PBE and (b) B3LYP references.  One curve per nested auxiliary set,
    each taken over the 29 excitation energies of that reference, with the
    optimum of each marked.  All values are RMSE in eV.  The $s$ and $sp$
    scans stop at $\theta = 1.00$.}
  \label{fig:defects}
\end{figure}

The nested shell sets converge and the result is shown in Figure~\ref{fig:defects}.  Over the same
systems and references, and each evaluated at its own optimum, the
root-mean-square deviation is $0.089$~eV with the $s$ set at $\theta = 0.65$,
$0.033$~eV with $sp$ at $0.55$, and $0.0054$~eV with $spd$ at $1.15$ -- each
added shell reducing the error by about a factor of three and then six.  This
is the clearest demonstration in this work that the auxiliary-set size acts as
a convergence parameter.  The two references behave almost identically: taken
separately they give $0.081$, $0.030$ and $0.0050$~eV for PBE and $0.095$,
$0.036$ and $0.0055$~eV for B3LYP, with the $spd$ optimum at $\theta = 1.20$
and $1.25$ respectively, so neither the convergence nor the recommended
$\theta$ depends on the choice of functional.

A single $\theta$ serves all seven defects.  Six of
the individual optima lie between $0.90$ and $1.40$ (Table~S3); the silicon--vacancy
center has two nearly equal minima, near $0.25$ and near $1.00$, which differ
by $15\%$ and the higher of which lies in the same range as the rest.  Adopting
the shared value of $1.15$ raises the error of the worst-affected system, the
$\mathrm{C_B C_N}$ pair, from $0.0029$ to $0.0053$~eV, and every defect remains
below $0.012$~eV.  The combined minimum is flat between $\theta = 1.10$ and
$1.25$.  That one scale factor should serve systems built from carbon, boron,
nitrogen, oxygen, magnesium and silicon is consistent with the tolerance
argument of Section~\ref{sec:metal_complexes}: these supercells carry large
auxiliary sets, and the tolerance in $\theta$ grows with them.  At the shared
optimum the carbon exponent is $\alpha_{\mathrm{C}} = 0.759$~bohr$^{-2}$,
against $0.726$ for the zinc complex and $0.726$ for the Thiel set, both also
with $spd$.

\subsection{Guidelines for Choosing $\theta$}
\label{sec:guidelines}

In this section we give practical guidance on choosing $\theta$ for a new system: which auxiliary set to use, what transfers between systems and what does not, and how to check a transferred value without a reference calculation.

First, the optimum moves with the shell set, so a
$\theta$ taken from one set does not apply to another: adding the $d$ functions
raises $\alpha_{\mathrm{C}}$ from $0.429$ to $0.726$~bohr$^{-2}$ and lowers the
error on the Thiel set by a factor of six.  The same ordering holds for the
point defects, where each added shell reduces the error by about a factor of
three and then six.  We therefore use $spd$ throughout, and the guidance below
applies to that set.

Within the same element the exponent transfers.  $\mathrm{Au_2}$ fixes
$\alpha_{\mathrm{Au}} = 0.204$~bohr$^{-2}$, and carrying that value to the
286-atom $\mathrm{Au_{25}}$ dimer reproduces conventional \pprpa{} to within
the convergence threshold of the reference calculation.  To test whether this
is a property of gold or of the method, we repeated the exercise on a second
metal.  $\mathrm{Cu_2}$ fixes $\alpha_{\mathrm{Cu}} = 0.245$~bohr$^{-2}$ at a
sharply defined optimum, and carrying that value to the 36-atom
$\mathrm{Cu_2(OAc)_4(H_2O)_2}$ dimer costs nothing measurable: the error of the
complex varies by half a per cent across $\theta = 2.40$ to $2.80$, a range
that contains the transferred value.

Between elements the exponent transfers less well, and whether that matters is
again decided by the size of the auxiliary set.  Carrying
$\alpha_{\mathrm{Au}}$ to $\mathrm{Ag_2}$, which has 18 auxiliary functions and
a sharp minimum, roughly doubles the error.  Yet a single $\theta$ serves all
seven point defects of Section~\ref{sec:periodic} -- systems built from carbon,
boron, nitrogen, oxygen, magnesium and silicon -- with every defect below
$0.012$~eV, because those supercells carry large auxiliary sets and
correspondingly flat minima.

We therefore recommend calibrating $\theta$ on the smallest system containing
the element that carries the excitation, and transferring $\alpha_A$ rather
than $\theta$.  Where calibration is impossible, $\alpha_A \approx
0.21$~bohr$^{-2}$ is appropriate for metal-centered excitations on a coinage
metal and $\alpha_{\mathrm{C}} \approx 0.73$--$0.79$~bohr$^{-2}$ for
ligand-centered and main-group systems with the $spd$ set.  The choice can be
tested without a reference calculation by recomputing at $\theta$ and
$\theta \pm 30\%$: a result insensitive to that variation lies on the flat
minimum characteristic of a large auxiliary set, and one that is not calls for
an explicit calibration.

\subsection{Computational Cost}
\label{sec:computational_time}

\begin{table}[htbp]
  \centering
  \caption{Wall-clock cost of conventional density-fitted \pprpa{} (noted here
    as conv.) and \pprpa{}-AS (noted here as aspd).  The integral
    transformation is the construction of the two-electron integral
    $\bm{L}$.  The eigenvalue solution is summed over the singlet and triplet
    manifolds, and is obtained by Davidson diagonalization except for the
    $\mathrm{Au_{25}}$ dimer, whose active space is small enough to
    diagonalize directly.  All finite-system calculations run on 32 threads.
    Times in seconds.}
  \label{tab:timing}
  \begin{tabular}{lrrrrrrrr}
    \toprule
    & \multicolumn{2}{c}{$N_{\mathrm{aux}}$}
    & \multicolumn{2}{c}{integral transformation}
    & \multicolumn{2}{c}{eigenvalue solution}
    & \multicolumn{2}{c}{total} \\
    \cmidrule(lr){2-3}\cmidrule(lr){4-5}\cmidrule(lr){6-7}\cmidrule(lr){8-9}
    System & conv. & aspd & conv. & aspd & conv. & aspd & conv. & aspd \\
    \midrule
    Au$_2$                    &   494 &   18 &   0.02 &   0.00 &   1.68 &  1.33 &   1.70 &   1.33 \\
    (3,6-SQ)$_2$Zn(Py)        &  3934 &  396 &  19.76 &   1.82 &   2.80 &  1.32 &  22.56 &   3.14 \\
    $[$Au$_2$(dppm)$_2]^{2+}$ &  5548 &  548 &  48.47 &   4.72 &   4.91 &  1.84 &  53.38 &   6.55 \\
    Au$_{25}$ dimer           & 11450 & 1598 & 889    & 122    &  38.3  & 36.5  & 927    & 158    \\
    NV in diamond\textsuperscript{\emph{a}}
                              &  3528 &  567 &  14.40 &   2.03 & 426.9  & 69.5  & 441.3  &  71.5  \\
    \bottomrule
  \end{tabular}

  \vspace{2pt}
  \raggedright\footnotesize
  \textsuperscript{\emph{a}} 63-atom supercell, PBE reference, 48 threads.
  Unlike the molecular rows, the two legs ran in separate jobs on the same
  partition and thread count rather than within one job.  Neither leg includes
  the cost of building its fitting object.
\end{table}

The approximation reduces the auxiliary basis by $84$ to $96\%$ and the total
cost by factors of $1.3$ to $8.2$ (Table~\ref{tab:timing}).  Both stages
contribute.  The integral transformation falls with $N_{\mathrm{aux}}$ in every
row.  The eigenvalue solution falls wherever it is iterative, because each
matrix--vector product contracts the auxiliary index.  It falls by $1.3$ for
$\mathrm{Au_2}$, $2.1$ for the zinc complex and $2.7$ for the digold complex.
The $\mathrm{Au_{25}}$ dimer is the exception.  Its $40/40$ space is small
enough to diagonalize directly, and a direct diagonalization does not see the
auxiliary basis once the matrix is built, so its solution moves only from
$38.3$ to $36.5$~s.  Which stage dominates sets the overall gain.  The
transformation takes $1.2\%$ of the runtime for $\mathrm{Au_2}$, which
therefore gains almost nothing, and $95.9\%$ for the $\mathrm{Au_{25}}$ dimer.

The nitrogen--vacancy center is the case the cost analysis of
Section~\ref{sec:pprpa_theory} describes.  Only $3.3\%$ of its runtime is spent
in the transformation.  The total still falls by a factor of six.  Its
$200/200$ active space is large enough that the matrix--vector product
dominates the solution.  The Davidson step is then
$O(N_{\mathrm{aux}} N_{\mathrm{vir}}^{3})$ and falls in proportion to
$N_{\mathrm{aux}}$, from $427$ to $69$~s.  The saving is largest where it was
intended to be.  Iterative solutions over large active spaces are what any
system needing this approximation will require.

These are excited-state times.  The ground-state calculation is unchanged and
is not included.  The charged-reference SCF remains, and so does the
construction of the conventional fitting object where one is needed.  For the
nitrogen--vacancy center that tensor occupies $11$~GB on disk.  The
approximation removes it from the excited-state calculation, but a reference
SCF is still required.  The saving applies to the part of the workflow that
scales worst with the number of states requested, not to the whole of it.

The natural next step is an analytical gradient.  The corresponding development
for TDDFT showed that the saving obtained for energies does not carry over
unchanged.  Terms absent from the energy expression reappear in the derivative,
and the cost advantage is
smaller.\cite{Havenridge2023Analytical,Wang2024Implementation,Samarasinghe2026Implementation}
The same is expected here.  Two features of the present scheme are favorable.
The auxiliary exponents depend only on the element and on $\theta$, not on the
nuclear coordinates.  The derivative of the auxiliary basis therefore involves
the motion of the Gaussian centers alone, and contributes no
exponent-derivative terms.  And $\theta$ is fixed rather than variationally
optimized, so it generates no additional response equations.  The limiting cost
is expected to lie in the terms that carry no auxiliary index, in particular
the relaxed density, which this approximation does not touch.  The benchmarks
here already pair the reduced auxiliary basis with a reduced orbital space, as
in active-space \pprpa{}.\cite{Li2023Linear}  Choosing that space without a
prior calculation remains open, and we intend to pursue it.

\section{Conclusions}
In this work we have extended the minimal auxiliary basis to the
particle--particle random phase approximation, giving a method we call
\pprpa{}--AS.  One uncontracted Gaussian per shell per atom replaces the
conventional set.  The Coulomb metric is retained, so the scheme
remains a resolution of the identity, and $N_{\mathrm{aux}}$ follows from
atomic composition rather than the orbital basis.

We have done benchmark test on a variety of systems.  Across 32 molecules and clusters
and seven point defects,including organic chromophores, transition-metal complexes, a
286-atom nanocluster and four crystalline hosts. The auxiliary basis shrinks
by $84$ to $96\%$, and conventional \pprpa{} excitation energies are reproduced
to within $0.03$~eV throughout.  The largest errors belong to the smallest
systems.  Every system beyond the organic benchmark set and the bare gold dimer
is reproduced to better than $0.011$~eV, and for the $\mathrm{Au_{25}}$ dimer
the difference falls below the convergence threshold of the reference
calculation itself. 

The computational costs saving is largest where it is needed.  Small molecules gain little and do
not need the approximation; the nitrogen--vacancy center gains sixfold, because
its iterative solve scales with $N_{\mathrm{aux}}$.  The tolerance in $\theta$
widens with the auxiliary set, so an exponent calibrated on a small analogue
carries into the large system it is meant for.  The $\mathrm{Au_{25}}$ dimer
shows what follows: its open-shell singlet and triplet come from one
closed-shell reference, without symmetry breaking or spin projection, at a
scale where conventional density fitting is the limiting cost. In the next step, we plan to derive the analytical gradients for this method.

\section*{Supporting Information}

Supporting information provides per-molecule errors for the Thiel set with the $spd$ auxiliary set; errors for
the three metal-containing systems under four reference determinants, with the
optimal scaling factor for each; fragment localization of the frontier orbitals
of the $\mathrm{Au_{25}}$ dimer dianion; and per-defect errors for the seven
point defects, at their individual optima and at the shared scaling factor
used in the main text.

\section*{Acknowledgments}

This work was supported by the Innovation Program for Quantum Science and Technology (Grant No. 2021ZD0303306), the National Natural Science Foundation of China (Grant No. 22393913), and the Strategic Priority Research Program of the Chinese Academy of Sciences (Grant No. XDB0450101). Authors gratefully acknowledge Chaoqun Zhang for insightful discussions of \pprpa{}.

\bibliography{reference}

\end{document}